\documentclass{ieeetj}
\usepackage{cite}
\usepackage{amsmath,amssymb,amsfonts}
\usepackage{algorithmic}
\usepackage{graphicx,color}
\usepackage{textcomp}
\usepackage{xcolor}
\usepackage{hyperref}
\hypersetup{hidelinks=true}
\usepackage{textcomp}
\let\labelindent\relax
\usepackage{enumitem}
\let\labelindent\IEEElabelindent
\usepackage{array}      
\usepackage[normalem]{ulem}
\usepackage{tabularx}   
\usepackage{booktabs}   
\usepackage{xcolor}
\usepackage{hyperref}
\usepackage{algorithm,algorithmic}
\def\BibTeX{{\rm B\kern-.05em{\sc i\kern-.025em b}\kern-.08em
    T\kern-.1667em\lower.7ex\hbox{E}\kern-.125emX}}
\AtBeginDocument{\definecolor{tmlcncolor}{cmyk}{0.93,0.59,0.15,0.02}\definecolor{NavyBlue}{RGB}{0,86,125}}

\def\OJlogo{\vspace{-4pt}$<$Society logo(s) and publication title will appear here.$>$}
\def\seclogo{\vspace{10pt}$<$Society logo(s) and publication title will appear here.$>$}

\def\authorrefmark#1{\ensuremath{^{\textbf{#1}}}}

\begin{document}
\receiveddate{XX Month, XXXX}
\reviseddate{XX Month, XXXX}
\accepteddate{XX Month, XXXX}
\publisheddate{XX Month, XXXX}
\currentdate{XX Month, XXXX}
\doiinfo{XXXX.2022.1234567}

\markboth{}{Author {et al.}}

\title{Towards Signal Processing In Networks: Multiplierless Block Transforms in Data Plane}

\author{Shota Ono\authorrefmark{1},
Marika Morikawa\authorrefmark{2},
Kazuma Kaneko\authorrefmark{2},
Junya Hara\authorrefmark{1},
Hiroshi Higashi\authorrefmark{3},
Shigenori Tani\authorrefmark{2}, and
Yuichi Tanaka\authorrefmark{1}
\affil{
\textsuperscript{}School/Graduate School of Engineering, The University of Osaka, Japan
\quad
\mbox{\textsuperscript{2}Information Technology R\&D Center, Mitsubishi Electric Corporation, Japan}
\quad
\textsuperscript{3}Faculty of Engineering Science, Kansai University, Japan}
\corresp{Corresponding author: Shota Ono (email:s.ono@msp-lab.org)}
}


\begin{abstract}
In this paper, we propose a signal processing concept, signal processing in networks~(SPIN), that executes signal processing tasks directly on programmable switches. 
Traditionally, signal processing tasks, such as denoising, decomposition, and dimensionality reduction, are performed on edge or cloud servers, i.e., raw streams are transmitted from sensors. This imposes a heavy processing load on edge/cloud servers and requires substantial bandwidth.
SPIN addresses this problem by performing signal processing on the forwarding path.
As an initial realization of SPIN, we  implement a DCT-based block transform on a programmable switch. We redesign its computation using a lifting-based lattice structure, multiplierless approximation, and pipeline partitioning across stages in a programmable switch.
We verify at the compiler level that the proposed implementation satisfies the strict hardware constraints of the Intel Tofino2—specifically regarding pipeline stages, packet header vector, and ternary content-addressable memory usage—and therefore enables line-rate processing. We further quantify the per-packet latency of the ingress match-action pipeline.
The SPIN-oriented preprocessing adds $46$~cycles (approximately $31$~ns at $1.5$~GHz). The total ingress match-action unit latency is approximately $87$~ns, which remains consistent with the low-latency operation targeted by SPIN.
Furthermore, we implement the proposed block transform on a BMv2 switch using P4 and evaluate it as preprocessing for in-network machine learning. Experimental results demonstrate that the proposed method improves classification performance over unfiltered baselines. 
\end{abstract}



\maketitle

\section{INTRODUCTION}
\IEEEPARstart{M}{any} signal processing systems, such as IoT monitoring, environmental sensing, biomedical wearables, and connected vehicles, deploy massive numbers of sensors, which continuously generate large-scale data streams~\cite{saleem2021deep, brzychczy2025process, anikwe2022mobile, lim2021state, baek2025edge}. These streams often contain noise, redundant samples, and task-irrelevant frequency components. 
Downstream tasks therefore require denoising, subband decomposition, dimensionality reduction, and feature extraction~\cite{s20216076}.

A straightforward approach to this processing is to transmit raw streams to cloud or edge servers
and perform signal processing at the endpoints of the network path~\cite{shi2016edge, liang2024review}. These
servers offer rich computing resources and high reconfigurability. However,
under this approach, the full raw streams must traverse the network. This consumes network bandwidth and increases end-to-end latency. Moreover, the processing load on these servers scales with the number of sensors, which limits scalability. This trade-off between processing flexibility and communication cost raises a central design question: Where should signal processing be performed?

\begin{figure}[t]
  \centering
  \includegraphics[width=1.0\linewidth]{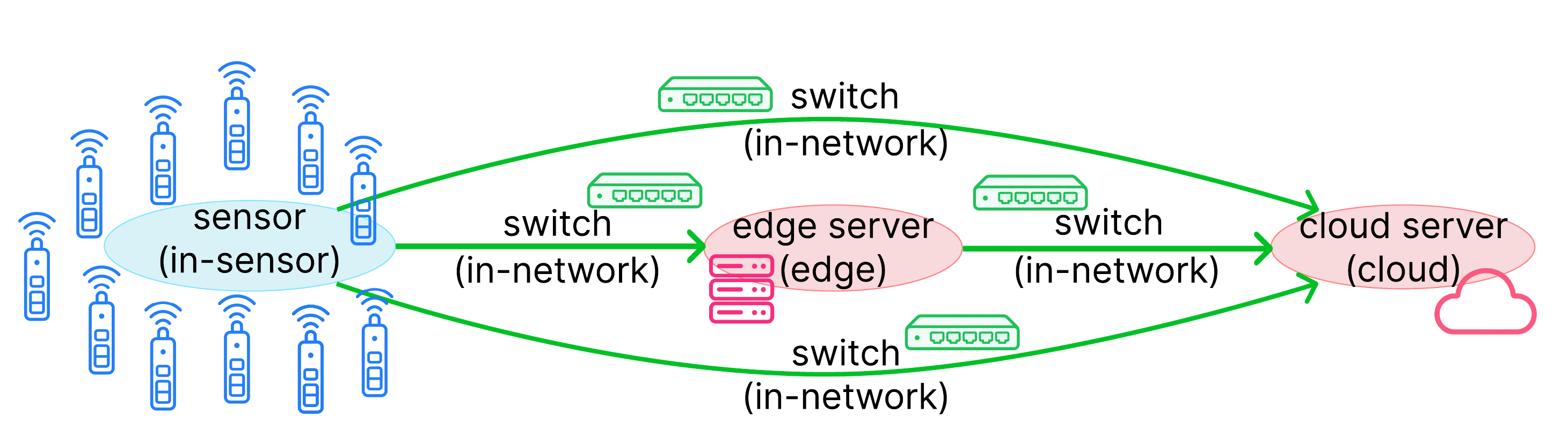}
  \caption{Computing paradigms along the path from sensors to the cloud.
Computation can be placed at sensors (in-sensor computing), on-path
switches (in-network computing), or downstream servers (edge and cloud
computing).}
  \label{fig:hierarchy}
\end{figure}

\begin{table*}[t]
\caption{Qualitative comparison of representative computing paradigms
for sensor data streams.}
\label{tab:tradeoff_comparison}
\centering
\footnotesize
\setlength{\tabcolsep}{4pt}
\renewcommand{\arraystretch}{1.15}
\begin{tabularx}{\textwidth}{@{}
>{\raggedright\arraybackslash}p{3.0cm}
>{\raggedright\arraybackslash}p{3.6cm}
*{2}{>{\raggedright\arraybackslash}X}
l
@{}}
\toprule
\textbf{Paradigm} & \textbf{Computing substrate} &
\textbf{Raw-data transport} &
\textbf{Processing flexibility} &
\textbf{Per-packet latency overhead} \\
\midrule
Cloud computing & Data-center servers & source (or edge) to cloud & Very high & High \\
Edge computing & Servers near data sources & source to edge & High & Medium \\
In-network computing & Programmable switches & source to switch & Medium & Very low \\
In-sensor computing & Sensor-integrated circuits & None & Low & None \\
\bottomrule
\end{tabularx}
\end{table*}
Several computing paradigms place computation closer to the data sources, as shown in Fig.~\ref{fig:hierarchy}, to reduce the cost of transmitting large data streams. These paradigms reduce both the amount of data transmitted over the network and the processing load imposed on downstream servers. 
Representative examples are in-sensor computing, in-network computing,
and edge computing. In-sensor computing reduces data movement at the
source. However, the available computation is constrained by the
sensor-side hardware~\cite{zhou2020near, baek2025edge}. 
Edge computing
provides higher processing flexibility, but raw streams are still transmitted over
the network to the edge server. As a complementary approach, \textit{in-network computing} performs lightweight
computation between sensors and edge servers (and/or edge and cloud servers), typically on programmable switches along the forwarding path. These
operations, however, must satisfy strict architectural
constraints~\cite{bosshart2013forwarding, ports2019should,
kianpisheh2022survey}. Table~\ref{tab:tradeoff_comparison} summarizes
these trade-offs.

Existing studies related to in-network computing
mainly target network-centric functions or task-specific
inference~\cite{kianpisheh2022survey}, such as
caching~\cite{jin2017netcache}, DDoS mitigation~\cite{zhang2020poseidon},
heavy-hitter detection~\cite{singh2022hh}, and in-network machine
learning~(INML)~\cite{xiong2019switches, planter}.
In contrast, this work considers observed sensor streams that may contain noise,
redundant samples, and unnecessary frequency components, which can
degrade the performance of such in-network tasks.
Closing this gap requires signal processing on the forwarding
path itself, before the streams reach the in-network logic or
downstream nodes.
 
In this paper, we propose \textit{signal processing in networks}~(SPIN), a concept for performing signal processing directly on programmable switches along the forwarding path. 
The goal of SPIN is to complement cloud- or edge-side computing by performing lightweight signal processing on the packet path before data reaches downstream computing nodes.\footnote{Note that SPIN differs in objective from classical network coding, which combines messages at intermediate nodes to improve communication efficiency while enabling their recovery at destinations~\cite{ahlswede2000network, li2003linear}. SPIN instead applies signal processing operators, such as filtering and spectral decomposition, to sensor measurements for downstream tasks. Its focus is on realizing these operators within the arithmetic and resource constraints of programmable switches.}

As a first step toward SPIN, this paper focuses on realizing a block transform in the data plane. A block transform is a critically sampled FIR filter bank whose filter length equals the block size, and it is one of the fundamental building blocks in signal processing~\cite{vaidyanathan2002multirate, vaidyanathan2003theory, vetterli1987theory}. Block transforms are widely used for spectral decomposition, subband analysis, and preprocessing. Unfortunately, direct mapping of a conventional block transform implementation onto programmable switches is difficult because of the constraints imposed by the reconfigurable match table~(RMT) architecture underlying programmable switches~\cite{bosshart2013forwarding}. In particular, RMT in programmable switches only provides a feed-forward pipeline with a limited number of stages, limited per-stage arithmetic capability, and restricted data dependencies across stages. 

To address this challenge, we redesign the block transform computation so that it conforms to the hardware constraints of a programmable switch. Specifically, we leverage a lifting-based lattice structure to express filtering as a cascade of elementary lifting operations, adopt an algorithmic approximation to eliminate multiplications, and split the overall computation across pipeline stages to preserve the required data flow. Through this redesign, we show that the signal processing system can be embedded into the switch pipeline while satisfying the hardware constraints of the programmable switch.

The experimental results demonstrate two findings. 
First, the proposed implementation satisfies the strict hardware constraints of the Intel
Tofino2~\cite{tofino2} at the compiler level. 
The implementation therefore runs at line rate, with an added ingress match-action unit~(MAU) latency of approximately 31 ns and a total ingress MAU latency of approximately 87 ns under a 1.5-GHz pipeline clock assumption.
Second, the proposed block transform enhances the classification
performance of subsequent INML tasks. 
These findings underscore both the feasibility 
and the effectiveness of performing signal processing in the data plane.
Finally, we discuss the potential for extending SPIN to IIR filters as a
future direction.


The remainder of this paper is organized as follows. 
Section~II introduces the RMT architecture and its constraints together with the binDCT used in our design. Section~III presents SPIN and the proposed multiplierless block transform, including its lattice reformulation and pipeline partitioning. Section~IV evaluates the hardware feasibility and effectiveness of the approach and discusses its extension to IIR filtering. Section~V concludes the paper.

\section{Preliminaries}

In this section, we describe the RMT architecture, the discrete cosine transform~(DCT)~\cite{rao1990discrete}, and the lifting-based multiplierless approximation of the DCT known as the binDCT~\cite{bindct}.

\subsection{RMT Architecture and Its Constraints}
RMT is a hardware architecture for processing packets in programmable switches~\cite{bosshart2013forwarding}. It is designed to achieve line-rate processing: Packets can be processed continuously without reducing the link speed. An RMT data plane consists of ingress and egress pipelines. The ingress pipeline processes packets before they enter the traffic manager, whereas the egress pipeline processes them after queueing. In this work, SPIN processing is implemented entirely in the ingress pipeline.

In RMT, packet processing is implemented as a multi-stage feed-forward pipeline, where each stage performs a part of the overall computation and forwards the intermediate result to the next stage.
Since each stage can execute only a limited set of operations~(e.g., additions/subtractions or bit shifts) within a single clock cycle, the entire computation must be partitioned across multiple stages. 

When signal processing is implemented on this architecture, three major challenges arise: Control-flow constraints, resource budget constraints, and arithmetic capability constraints. 

\noindent
\textbf{Control Flow.}
RMT employs a strictly feed-forward
architecture~\cite{bosshart2013forwarding}, which
prohibits loops and backward data flow across stages. 


\noindent
\textbf{Resource Budgets.}
A programmable switch imposes hardware-bounded resource budgets in the following areas.

\begin{itemize}
\item \textbf{Stage budget.}
The pipeline consists of a fixed number of stages. Each stage performs table matching and then executes the corresponding operations on packet data. Therefore, the entire computation must be partitioned across these stages in a feed-forward manner within the available stage budget.

\item \textbf{PHV capacity.}
Within the feed-forward pipeline, the packet header vector~(PHV) is the sole register
that carries data between stages for stateless
computation~\cite{bosshart2013forwarding, sivaraman2016packet}.
Its total bit width is hardware-bounded: A programmable switch
allocates only a fixed number of 8-, 16-, and 32-bit PHV
containers per pipeline. At every stage, the aggregate width of
variables that are \emph{live}---produced at or
before that stage and still required afterwards---must therefore
fit within this budget.

\item \textbf{Parallel arithmetic logic unit~(ALU) capacity.}
Each stage provides only a fixed number of ALU lanes, so the
number of ALU actions executed within a single stage cannot
exceed this lane count.
\end{itemize}

\noindent
\textbf{Arithmetic Capability.}
Each stage supports only a restricted instruction set, including addition, subtraction, and bit shifts. Floating-point arithmetic and variable multiplication/division are therefore unsupported. Combined operations such as shift-and-add on a single variable must be decomposed across multiple stages unless the underlying ALU natively supports the fusion~(see also Section IV-F). 

\vspace{0.5em}

The above three constraints collectively make direct implementation of conventional signal processing algorithms infeasible on programmable switches. Therefore, signal processing computations must be reformulated or approximated to reduce computational complexity.


\subsection{DCT and binDCT}
The DCT is a widely used orthogonal transform in signal processing, with applications including image compression~\cite{wallace1991jpeg}.

The binDCT~\cite{bindct}, which is derived from the lifting scheme, is a multiplierless approximation of the DCT.
It factorizes the DCT matrix into lifting steps and approximates the resulting irrational rotation parameters with dyadic fractions of the form $k/2^m$ ($k, m \in \mathbb{Z}$).
Since dyadic fractions are realizable by bit shifts and additions, all multiplications are eliminated.
For example, the eight-point binDCT computes the frequency-domain coefficients using only 30 additions and 13 bit shifts.
Figure~\ref{fig:binDCT_flow}(a) shows the signal-flow graph of the eight-point binDCT.

\section{SPIN and SPIN-oriented Multiplierless Block Transforms}
In this section, we present a switch-oriented realization of SPIN using multiplierless block transforms. We first position SPIN within existing signal processing paradigms. We then formulate the block transform for SPIN and introduce a binDCT-based multiplierless realization. Finally, we present resource-aware pipeline partitioning that maps the resulting signal-flow graph to switch stages.

\subsection{SPIN Standpoint}
As shown in Fig. 1 and Table~\ref{tab:tradeoff_comparison}, SPIN complements major signal processing approaches. Compared with cloud- or edge-oriented signal processing, which offers high flexibility at the cost of communication overhead and latency, SPIN performs lightweight signal processing at line rate on the forwarding path.
Relative to in-sensor processing, which minimizes data movement but lacks runtime reconfigurability, SPIN can exploit the reconfigurability provided by programmable switches.

Although SPIN can borrow hardware-efficient ideas from VLSI-oriented DSP system design~\cite{bartholoma2008systematic}, such as lifting-based factorization and multiplierless approximation~\cite{bayer2013multiplierless}, the two differ fundamentally in their design freedom: VLSI implementations can tailor the datapath itself to the target algorithm, whereas SPIN must map signal processing onto the fixed RMT pipeline of programmable switches without modifying the hardware.

In this work, we focus on how to reformulate signal processing so that it conforms to the architectural constraints of the RMT pipeline. The key challenge lies in reconsidering signal processing computations to satisfy the three constraints described in Section II-A.

\subsection{Block Transforms for SPIN}
Here, we formulate a block transform for SPIN. Let  $\mathbf{x} \in \mathbb{R}^{N}$ be an input block of $N$ consecutive samples and let $\mathbf{y} \in \mathbb{R}^{K}$ be the  output block of $K$ coefficients. After a block transform, $\mathbf{y}$ is represented as follows.
\begin{equation}
    \mathbf{y} = \mathbf{H}\mathbf{x},
    \label{eq:block_transform}
\end{equation}
where $\mathbf{H} \in \mathbb{R}^{K \times N}$ is the block transform matrix. The $k$-th row of $\mathbf{H}$ corresponds to an analysis filter.

When $\mathbf{H}$ is dense, a direct implementation of this matrix-vector
product in the data plane is impractical under the RMT architecture constraints.
This is because such an implementation requires $O(NK)$ multiplications, while the RMT pipeline does not support general multiplication and has limited per-stage arithmetic capacity. 

Our key idea is to approximate $\mathbf{H}$
by a product of sparse, switch-executable operators:
\[
    \mathbf{H} \approx \mathbf{A}_{M-1} \mathbf{A}_{M-2} \cdots \mathbf{A}_0,
\]
where \[
\mathbf{A}_i \in
\mathbb{R}^{R_{i+1} \times R_{i}}
\]
denotes a switch-executable operator at stage $i$, and $R_{i}$ and $R_{i+1}$ denote the input and output dimensions, respectively. Each $\mathbf{A}_i$ must be implementable by operations supported in the data plane, such as addition, subtraction, and bit shifts. Under this formulation, the block transform is executed as a feed-forward sequence of elementary operations rather than as direct convolution or dense matrix multiplication.

For an analysis block transform such as the DCT, this type of factorization is
well established: fast DCT algorithms factorize the DCT matrix
into a cascade of butterfly operations and plane
rotations~\cite{chen1977fast, loeffler1989practical}. The
butterfly operations require only additions and subtractions, whereas
each rotation involves multiplications by irrational coefficients.
The binDCT (described in Section~II-B) resolves this
obstacle by expressing each rotation as a sequence of lifting steps with
dyadic-rational coefficients. We therefore adopt the binDCT
factorization as a concrete realization of this decomposition
: Butterfly
and lifting steps reduce to additions, subtractions, and
bit shifts, and the cascade realizes $\mathbf{H}$ without any general
multiplication. 

Moreover, the binDCT is an acyclic, feed-forward transform
and has no backward data dependencies. Hence, the choice of the algorithm already satisfies the control-flow constraint. The remaining two
constraints, the resource budget and the arithmetic capability, are
addressed by the pipeline partitioning described next.

\begin{figure}[t]
  \centering
  \includegraphics[width=1.0\linewidth]{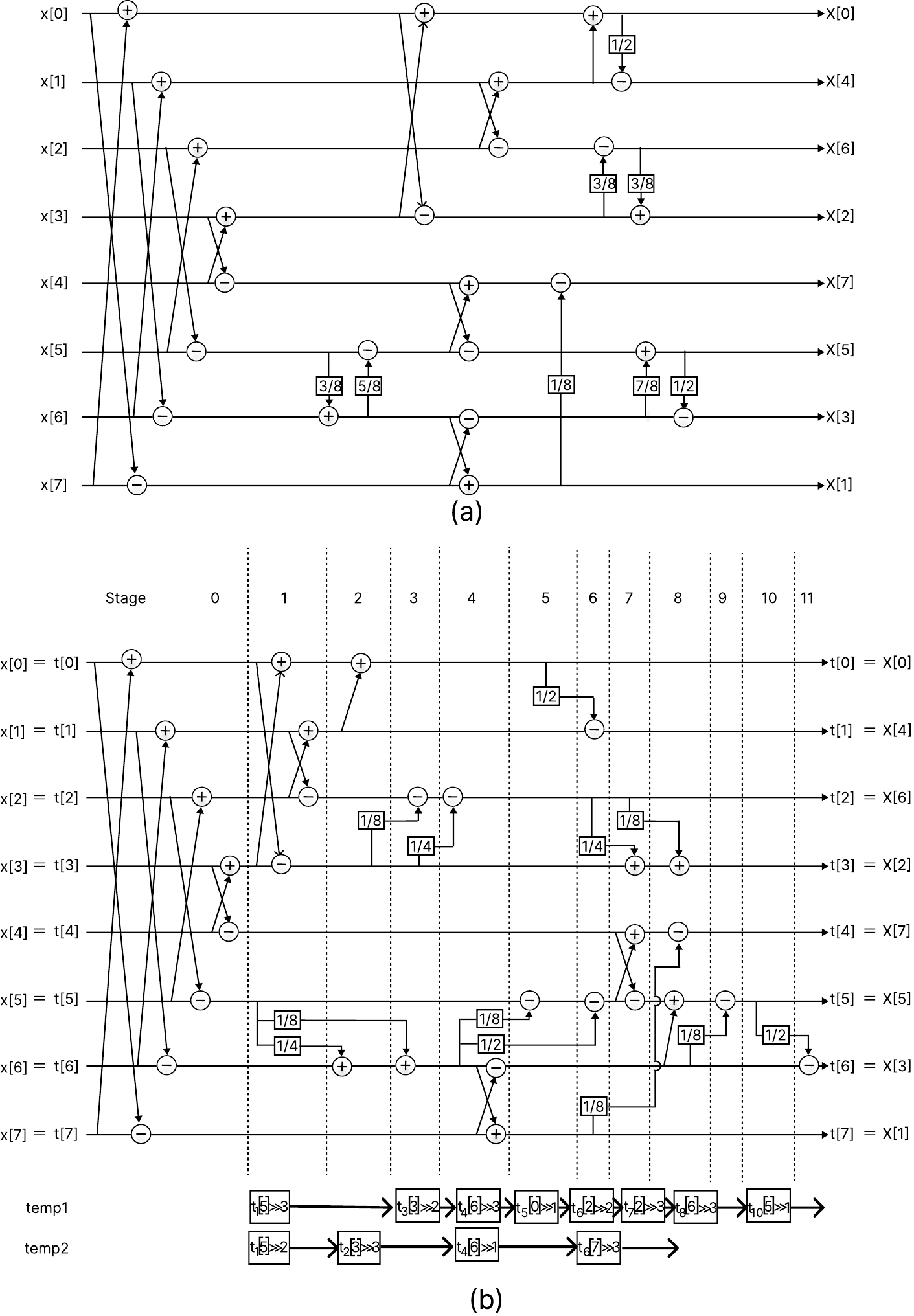}
  \caption{Signal-flow graphs of the eight-point binDCT and its redesigned
version for SPIN: (a) the original eight-point binDCT and (b) the
redesigned signal-flow graph for SPIN. In (b), the left labels $x[0]$--$x[7]$ are the eight samples of one input block, and the right labels $X[0]$--$X[7]$ are the resulting binDCT coefficients. The variables $t[0]$--$t[7]$ are working variables used throughout the computation. The two rows beneath the redesigned graph show the contents of the two temporary-variable slots (PHV containers \texttt{temp1} and \texttt{temp2}) at each stage, where \(t_i[j] \gg m\) denotes the value of the $j$th branch at stage $i$, shifted right by $m$ bits.}
  \label{fig:binDCT_flow}
\end{figure}

\subsection{Pipeline Partitioning for Programmable Switches}
The binDCT factorization removes general multiplications from the block transform computation. However, this factorization is not sufficient to make the computation directly executable on the RMT pipeline.
The remaining challenge is to map the resulting operations onto the pipeline under its resource constraints. The binDCT signal-flow graph, shown in Fig.~\ref{fig:binDCT_flow}(a), must be expanded into elementary switch operations and then assigned to pipeline stages under the RMT architectural constraints. 



To address this challenge, we redesign the signal-flow graph and partition it into a stage-by-stage pipeline structure. We design the signal-flow graph subject to the following rules derived from the hardware constraints. Fig.~\ref{fig:binDCT_flow}(b) shows the resulting eight-point binDCT for SPIN:

\begin{enumerate}
  \item \textbf{Per-stage capacity (arithmetic capability constraint):}
        Each node in Fig.~\ref{fig:binDCT_flow}(a) labeled with a
        dyadic coefficient (e.g., $3/8$) is
        decomposed into elementary shifts and additions
        (e.g., $3/8 = 1/4 + 1/8$).
  \item \textbf{Bounded resource usage 
      (resource budget constraint):}
        Decomposing dyadic-rational coefficients into shifts and
        additions/subtractions increases the number of intermediate values.
        Each value must reside in a PHV container from the stage that
        produces it until its last use.
        In programmable switches, the number of simultaneously live temporary values is constrained by the PHV capacity. In our Tofino2 implementation, schedules with at most two live temporary values can be successfully compiled. We therefore maintain this two-value bound at every stage.

Fig.~\ref{fig:binDCT_flow}(b) shows the redesigned signal-flow graph
together with the occupancy of the two temporary-variable slots across
pipeline stages. The two rows beneath the graph indicate the values
stored in these slots at each stage.
Here, $t_i[j]$ denotes the value of the $j$th branch at stage $i$, and
$t_i[j]\gg m$ denotes this value shifted right by $m$ bits. 

For example, at stage~1, both temporary variable slots are occupied by
the two shifted copies of $t_1[5]$, namely $t_1[5] \gg 3$ and
$t_1[5] \gg 2$, which together realize the $3/8$ coefficient as
$1/4 + 1/8$. At stage~2, $t_1[5] \gg 2$ is used for the last time, so its temporary variable slot becomes available and is reused to store $t_2[3] \gg 3$. Thus, the two slots at stage~2 contain
$t_1[5] \gg 3$ and $t_2[3] \gg 3$. Then, at stage~3,
$t_1[5] \gg 3$ used for the last time, and its slot is reused for
$t_3[3] \gg 2$. Consequently, the two slots at stage~3 contain
$t_2[3] \gg 3$ and $t_3[3] \gg 2$. In this way, at most two intermediate values are retained
simultaneously throughout the schedule.
\end{enumerate}

Note that this redesign changes only the execution schedule of the
elementary operations, not the transform itself; the redesigned binDCT is
functionally equivalent to the original binDCT in
Fig.~\ref{fig:binDCT_flow}(a). The two differ only in the rounding of
intermediate values, since decomposing a dyadic coefficient into separate
shift-and-add operations changes where truncation occurs in fixed-point
arithmetic.


\begin{table}[t]
    \centering
    \caption{Number of stages required per frequency component}
    \label{tab:performance_comparison}
    \resizebox{\columnwidth}{!}{
        \begin{tabular}{lcccccccc}
            \toprule
            Element & X[0] & X[1] & X[2] & X[3] & X[4] & X[5] & X[6] & X[7] \\
            \midrule
            Required Stages & 4 & 5 & 9 & 12 & 7 & 10 & 5 & 9 \\
            \bottomrule
        \end{tabular}
    }
\end{table}

\begin{table}[t]
    \centering
    \caption{Comparison of frequency-component configurations}
    \label{tab:config_comparison}
    \resizebox{\columnwidth}{!}{%
        \begin{tabular}{lccc}
            \toprule
            Configuration & Frequency components & Selection principle & Stages \\
            \midrule
            Baseline (Full)  & $X[0]$--$X[7]$              & All coefficients   & 12 \\
            Low-frequency    & $X[0],X[1],X[2],X[3]$      & Energy compaction    & 12 \\
            Selected         & $X[1],X[2],X[5],X[6],X[7]$ & Discriminative power & 10 \\
            \bottomrule
        \end{tabular}
    }
\end{table}

Table~\ref{tab:performance_comparison} presents the number of pipeline stages required to calculate each frequency-domain coefficient. By partitioning the computation across pipeline stages in this manner, the programmable switch can execute the processing at line rate.

\section{Evaluation}
In this section, we evaluate the proposed SPIN realization from two complementary perspectives: \emph{feasibility} and \emph{effectiveness}. Feasibility is assessed by compilation, resource usage, and per-packet latency, while effectiveness is measured by INML classification performance. 

We first describe the dataset and the testbed setup, including the bit-shifting scheme, hyperparameter selection, implementation, and evaluation metrics. We then report the results from the two perspectives: an analysis of its hardware resource consumption and latency, followed by the impact of SPIN filtering on INML classification performance.

\subsection{Dataset} 
To evaluate the effectiveness of the proposed SPIN filtering on INML, we use the Wafer dataset~\cite{olszewski2001generalized, 8894743}. This dataset comprises sensor measurement values recorded during the processing of silicon wafers for semiconductor manufacturing. 

Each input is a time-series signal with a length of 152 time steps, and the task is to classify the state as either normal or abnormal based on the waveform. Due to the fixed input/output length of 8 imposed by the binDCT, the original time-series signals of length 152 are segmented into subsequences of length $N=8$. Each subsequence is carried in one packet and classified independently by INML. Thus, the original sequence-level task is reformulated as packet-level classification.

Note that there is a significant class imbalance in this dataset; abnormal samples constitute only 10.7\% of the training data and 12.1\% of the test data. Accordingly, as described later, we account for this imbalance in both model training and performance evaluation.

\subsection{Testbed Setup}
To evaluate the effect of binDCT-transformed data on ML classification performance, we employ Planter~\cite{planter}, a framework designed to map pre-trained ML models onto a programmable switch.
This framework facilitates efficient experimental validation on the BMv2 target.

Planter provides several mapping methodologies that translate a trained tree-based model into the match-action pipeline. Among these, we adopt the Encode-Based~(EB) mapping. In EB mapping, a separate match-action table is used for each input feature. The table compares the feature value against the split thresholds extracted from the trained tree and encodes the resulting interval as a compact range identifier. The set of encoded feature values is then used as the key to a final decision table, which outputs the predicted label.

\subsubsection{Comparison Methods}
\begin{figure}[t]
  \centering
  \includegraphics[width=1.0\linewidth]{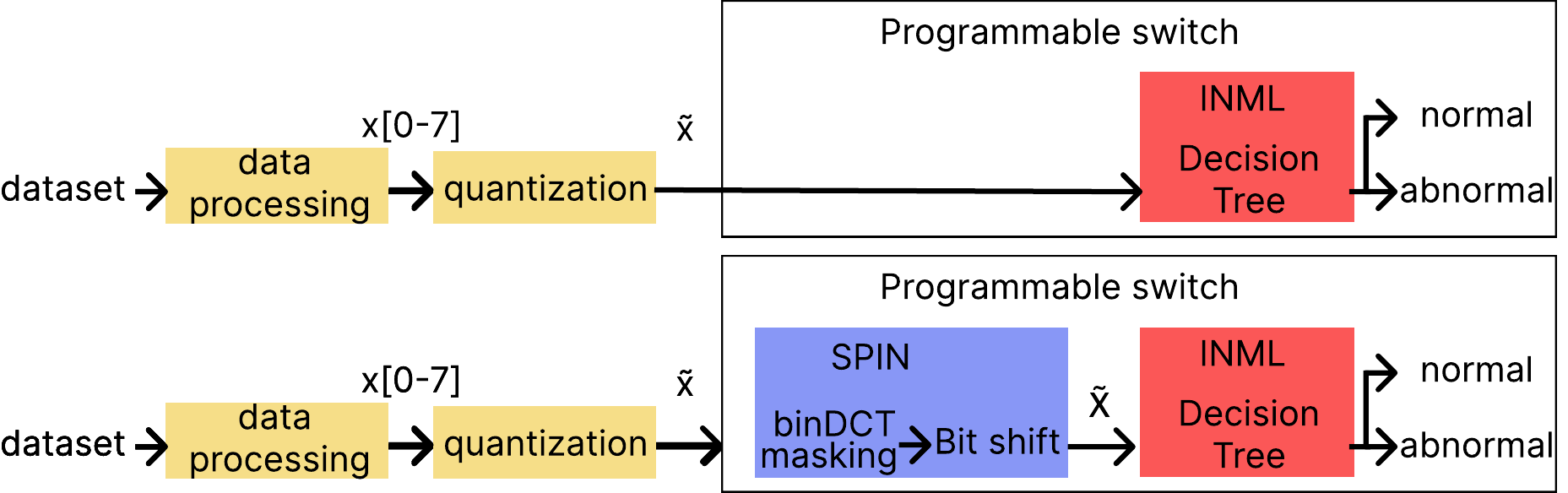}
  \caption{Processing pipelines for the raw-feature and binDCT-based settings.
The upper pipeline directly classifies quantized raw features with INML, whereas the lower pipeline applies binDCT masking and a data shift through SPIN before INML classification.
}
  \label{fig:dr}
\end{figure}

Figure~\ref{fig:dr} illustrates the data processing pipelines for the raw and the SPIN-based settings, respectively. Given a real-valued input vector $\mathbf{x}$ from the dataset, we apply preprocessing before classification.
We compare the following settings:

\begin{itemize}
    \item \textbf{Raw}: The input values are first quantized, and the quantized values are fed directly into the INML stage without applying binDCT.
    \item \textbf{binDCT}: The quantized values are transformed by the SPIN-oriented binDCT. The transformed coefficients are then shifted to satisfy the non-negative integer input constraint of Planter, and the resulting features are fed into the INML stage.
    \item \textbf{binDCT-based filtering}: A subset of the binDCT coefficients is zeroed out to perform filtering before the bit shift. The shifted features are then fed into the INML stage.
\end{itemize}


\subsubsection{Bit Shift}
Programmable switches do not support floating-point arithmetic, and Planter imposes a constraint that input values must be non-negative integers.
On the other hand, programmable switches can 
natively handle wide integers (e.g., 32-bit or 64-bit registers 
and metadata fields), so large-magnitude integer values pose no 
practical difficulty. 

We preprocess the raw data using the following level-shifting and quantization procedure before feeding it into the programmable switch:
\begin{equation}
    \tilde{\mathbf{x}} = \lfloor 2^{12}\, \mathbf{x}+ 2^{14}\mathbf{1}_{8} \rfloor, 
    \label{eq:input_preprocessing}
\end{equation}
where $\mathbf{x}$ is the original signal value and $\tilde{\mathbf{x}}$ is the input value to the programmable switch. This preprocessing converts the real-valued sensor input into a non-negative integer.

The INML classifier also requires non-negative integer inputs. The final stage of the binDCT pipeline therefore applies the following data shift before feeding the coefficient to INML:
\begin{equation}
    \tilde{\mathbf{X}} = \mathbf{X} + 2^{18}\mathbf{1}_{8}, \label{eq:level_shift}
\end{equation}
where $\mathbf{X}$ is the binDCT coefficient computed on the programmable switch, and $\tilde{\mathbf{X}}$ is the shifted SPIN output fed to INML.

To isolate the impact of the quantization in (\ref{eq:input_preprocessing}), we apply the binDCT on the server without quantization and evaluate the resulting ML classification performance. The server-side and switch-side configurations thus differ in their arithmetic: The server uses floating-point arithmetic, whereas the switch uses integer arithmetic.

Note also that our implementation computes only the forward binDCT and omits the IDCT stage, since the inverse transform would consume additional pipeline stages.








\subsubsection{Feature Selection and Hyperparameters}
\label{sec:hyperparameter}
The dataset is partitioned into training, validation, and testing sets, with the validation set used to optimize the configuration. The Wafer dataset provides predefined training and testing sets containing 1,000 and 6,164 samples, respectively. We split the predefined training set into training and validation sets at an 80:20 ratio. We employ a decision tree classifier for INML due to its robustness to the data shift described above, and the tree depth is restricted to six. 
To address class imbalance, we weight each class inversely to its frequency in the training set.

The remaining design choice concerns which binDCT coefficients are used for classification. 
For our experiments, we use the \emph{Selected} configuration that uses \(X[1], X[2], X[5], X[6], X[7]\) for INML based on the validation-set MCC. For the server, $X[1], X[3], X[4], X[5], X[6], X[7]$ are selected separately using the same criterion. We also use the \emph{Baseline} configuration that uses all coefficients \(X[0]\)--\(X[7]\).

Table~\ref{tab:performance_comparison} reports the number of pipeline stages
required for each binDCT coefficient.
While the Low-frequency configuration uses only four coefficients, it requires 12 pipeline stages—the same as the Baseline configuration—because computing X[3] involves the longest dependency path.
In contrast, the Selected configuration, which does not include X[3], requires only
\(10\) stages.

This behavior arises from the data dependencies in the binDCT signal-flow
graph shown in Fig.~\ref{fig:binDCT_flow}(b). The ordering of the butterfly
and lifting operations constrains their assignment to pipeline stages.
Consequently, the required pipeline depth is determined by the longest
dependency path among the retained coefficients rather than by the number of
retained coefficients.

\subsubsection{Evaluation Metrics} We evaluate SPIN from two perspectives---feasibility and effectiveness---using the following metrics.
\begin{enumerate} 
\item[(1)] \textbf{Resource Consumption (feasibility):} Memory utilization and the number of pipeline stages are measured to evaluate the feasibility of the implementation.
These hardware resource metrics are analyzed using the Open P4 Studio compiler~\cite{p4studio}, which targets the Tofino2 Native Architecture~(T2NA). 
The compiler rejects any program that exceeds the T2NA pipeline constraints, including stage count, PHV allocation, and TCAM capacity. We therefore treat successful compilation as evidence of feasibility.

\item[(2)] \textbf{Latency (feasibility):} We quantify the per-packet
processing latency introduced by SPIN in the ingress pipeline.
We define the latency of a configuration as the number of clock cycles a packet
spends in the ingress pipeline, obtained from the Open P4 Studio compiler. 
To report absolute time, we convert cycles to nanoseconds assuming a pipe clock of
approximately $1.5$~GHz, derived from the published Tofino2 aggregate rate of
$6$~Gpackets/s across four pipelines~\cite{agrawal2020intel, shirmarz2025network}. 

\item[(3)] \textbf{ML Performance (effectiveness):} We use the following five metrics, where cls$0$ denotes the abnormal~(anomaly) class, which is the minority class of interest:
\begin{itemize}
  \item \textbf{Matthews correlation coefficient~(MCC):} A single confusion-matrix-wide 
  correlation in $[-1, 1]$~\cite{chicco2020advantages}.
  \item \textbf{Area under the precision--recall curve~(PR-AUC)$_{\text{cls}0}$:} We use the area under the precision–recall curve for cls$0$. This metric is robust to the number of true negatives and is therefore suited to imbalanced classification~\cite{kim2022empirical}.
  \item \textbf{F1$_{\text{cls}0}$:} The F1 score for cls$0$, defined as
  \begin{equation}
    F_1 = \frac{2\,\mathrm{Precision}\cdot\mathrm{Recall}}
               {\mathrm{Precision}+\mathrm{Recall}}.
    \label{eq:f1}
  \end{equation}
  The F1 score is the harmonic mean of precision and recall and gives equal weight to both. We use it as a threshold-level summary of the balance between false positives and false negatives, while PR-AUC and MCC characterize ranking quality and class-imbalance-aware performance, respectively.
  \item \textbf{Precision$_{\text{cls}0}$, Recall$_{\text{cls}0}$:} 
  The class-wise precision and recall for cls$0$, reported separately 
  to show the trade-off underlying the aggregate metrics above.
\end{itemize}

We implement the filtering logic described in the previous section on the BMv2 software switch using Planter, and assess the impact of SPIN on classification performance. Because the INML implementation outputs only hard class labels, we compute PR-AUC on the server using the same trained decision tree and identical preprocessing as on the switch.
\end{enumerate}

\subsection{Feasibility}
We evaluate the resource consumption of the proposed SPIN filtering mechanism using Open P4 Studio. The detailed memory usage and pipeline stage requirements are summarized in Table~\ref{tab:resource_analysis}.

\begin{table}[t]
  \centering
  \caption{Resource Consumption Analysis of the Proposed System}
  \label{tab:resource_analysis}
  \begin{tabular}{lccc}
    \toprule
     & \textbf{raw} & \textbf{binDCT} & \textbf{filtered} \\
    \midrule
    \textbf{TCAM (units)} & 8 & 8 & 5 \\
    \textbf{Ingress PHV (bits)} & 560 & 624 & 616 \\
    \textbf{Stages} & 2 & 15 & 13 \\
    \bottomrule
  \end{tabular}
\end{table}

Table~\ref{tab:resource_analysis} reports the resource cost of binDCT-based filtering, taking INML~(raw)---which quantizes the input and feeds it
directly to the decision tree without binDCT---as the baseline. Three trends are observed.

First, TCAM usage decreases from $8$ to $5$ units: Filtering selects only the five frequency-domain coefficients as decision-tree features.

Second, the pipeline depth grows from $2$ to $13$ stages. These $13$ stages can be decomposed into $10$ for the binDCT computation, $1$ for the INML input data shift, and $2$ for decision-tree inference; the binDCT transform is therefore the dominant contributor. 

Third, PHV usage rises from $560$ to $616$~bits. The binDCT requires two 32-bit temporary variables to retain intermediate values across pipeline stages. 
These variables add 64 bits to the PHV allocation. At the same time, filtering reduces the number of INML features from eight to five. This reduction allows the compiler to use one fewer 8-bit PHV container. The net increase is therefore 56 bits.

Overall, binDCT-based filtering increases pipeline depth from $2$ to $13$
stages and PHV usage by $56$~bits, while reducing TCAM usage from $8$ to
$5$ units. This additional cost accompanies the classification gain reported in Section~IV-E;
the successful compilation with Open P4 Studio confirms that the resulting
design still fits within the hardware constraints.

We next isolate the effect of frequency-domain filtering by comparing
\emph{INML~(filtered binDCT)} with \emph{INML~(binDCT)}, i.e., the
Selected and Baseline configurations in Table~\ref{tab:config_comparison}.
Masking reduces all three resources: TCAM usage decreases from $8$ to
$5$ units, pipeline depth from $15$ to $13$ stages, and PHV usage from
$624$ to $616$~bits. These reductions follow from using only five of the eight binDCT coefficients as INML features.

\vspace{0.5em}

\subsection{Latency Analysis}
\begin{figure}[t]
  \centering
  \includegraphics[width=\linewidth]{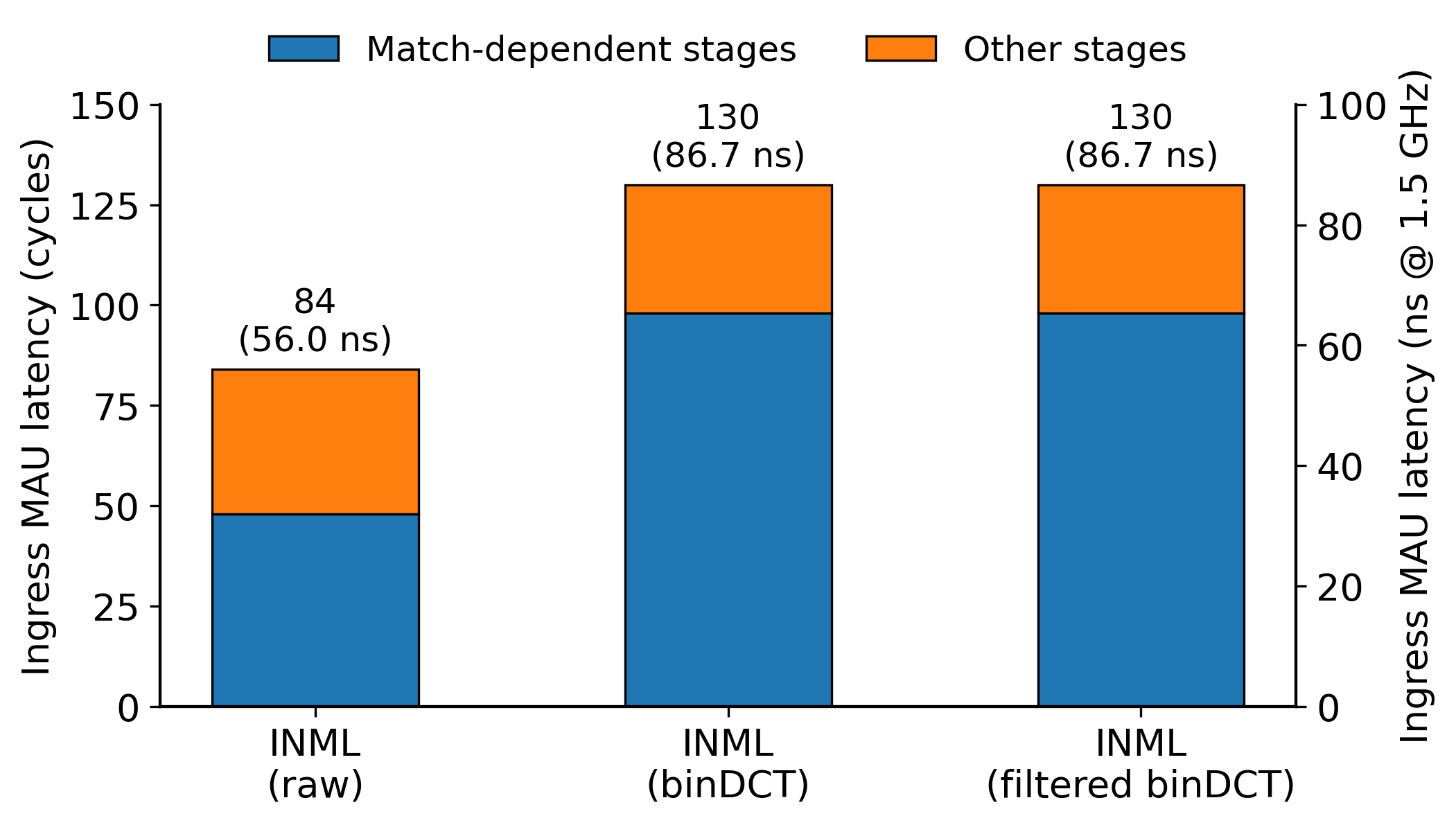}
  \caption{Ingress MAU latency of the three configurations, decomposed into
  match-dependent stages and other stages. The left axis is in clock cycles
  and the right axis is the corresponding wall-clock latency at the
  $1.5$-GHz MAU clock.}
  \label{fig:latency}
\end{figure}

Figure~\ref{fig:latency} reports the ingress MAU latency of the three configurations in clock cycles and in nanoseconds.
Taking INML~(raw) as the baseline, we observe two trends.

First, binDCT increases the latency from $84$ to $130$ cycles
($56.0$ to $86.7$~ns, assuming a $1.5$-GHz MAU clock). This increase is
caused by two additional match-dependent stages. 
These stages can begin their table lookups only after the previous stages have finished computing the corresponding match keys.
This situation occurs because an offset is added according to
(3) before the binDCT coefficients are used as inputs to INML. The compiler
places these offset additions in two stages, followed by the corresponding
INML lookup stages. As a result, the two lookup stages become match-dependent,
accounting for the $46$-cycle increase.

Second, coefficient masking reduces the pipeline depth from 15 to 13 stages, but a smaller pipeline depth does not necessarily translate into lower latency. In a match-action
stage, table matching is followed by action execution. INML (binDCT) and INML (filtered binDCT) have the same number
of match-dependent stages, and therefore both require $130$ cycles.

\begin{table}[t]
\centering
\caption{Classification performance on the Wafer dataset.}
\label{tab:Wafer_result}
\small
\setlength{\tabcolsep}{4pt}
\begin{tabular}{lccc@{\hskip 12pt}ccc}
\toprule
 &  \multicolumn{3}{c}{Server} &
  \multicolumn{3}{c}{INML} \\
\cmidrule(lr){2-4} \cmidrule(lr){5-7}
Metric & Raw & binDCT & Filtered & Raw & binDCT & Filtered \\
\midrule
PR-AUC$_{\text{cls}0}$    & 0.32 & 0.54 & \textbf{0.56} & 0.32 & 0.42 & \textbf{0.49} \\
MCC                       & 0.27 & 0.41 & \textbf{0.42} & 0.27 & 0.35 & \textbf{0.42} \\
F1$_{\text{cls}0}$        & 0.36 & 0.47 & \textbf{0.48} & 0.36 & 0.43 & \textbf{0.49} \\
Precision$_{\text{cls}0}$ & 0.28 & \textbf{0.38} & \textbf{0.38} & 0.28 & 0.35 & \textbf{0.46} \\
Recall$_{\text{cls}0}$    & 0.49 & 0.63 & \textbf{0.65} & 0.49 & \textbf{0.56} & 0.53 \\
\bottomrule
\end{tabular}
\end{table}

\subsection{Effectiveness}
The results are summarized in Table~\ref{tab:Wafer_result}. 
Table~\ref{tab:Wafer_result} demonstrates that binDCT-based filtering improves classification performance.
 
When the binDCT is applied without filtering, the switch-side implementation underperforms its server-side counterpart. The MCC decreases from 0.41 to 0.35 for INML~(binDCT). This gap is partly attributable to the input-level quantization  introduced by \eqref{eq:input_preprocessing}:
The server uses floating-point inputs, whereas the switch uses quantized integers whose error propagates through the binDCT.

BinDCT-based filtering substantially mitigates the performance degradation caused by quantization. Comparing the two switch-side configurations INML~(filtered binDCT) and INML~(binDCT),
filtering improves PR-AUC$_{\text{cls}0}$, MCC, F1$_{\text{cls}0}$, and Precision$_{\text{cls}0}$. 

Notably, the MCC obtained by INML~(filtered binDCT) ($0.42$)  matches the server-side value ($0.42$). This result indicates that zeroing the non-discriminative coefficients mitigates the performance degradation associated with switch-side integer processing, allowing the MCC to reach the level achieved in the server-side evaluation.

\subsection{Extension to an IIR Filter}
Extending SPIN to IIR
filters is more difficult under the RMT architectural constraints, since an IIR filter feeds previous
\emph{outputs} back into later updates. Therefore, the
data plane must preserve a filtered value across packets and use it when the
next packet reaches the pipeline.

Consider the first-order exponentially weighted moving average~(EWMA) filter with the following one-dimensional recurrence:
\begin{equation}
    x_{\text{filtered}}[n] =
    \begin{cases}
        x[n], & n = 0, \\
        \alpha\, x[n] + (1-\alpha)\, x_{\text{filtered}}[n-1], & n \ge 1,
    \end{cases}
    \label{eq:ewma}
\end{equation}
where $\alpha \in [0,1]$ is the smoothing coefficient.

Since the RMT pipeline provides no general multiplier, we restrict the
coefficient to $\alpha = 1 - 2^{-i}$ so that every product in
\eqref{eq:ewma} reduces to a right shift. Under this restriction, the
update can be rewritten as
\begin{equation}
x_{\text{filtered}}[n]
=
\bigl(x[n] - 2^{-i} x[n]\bigr)
+
2^{-i} x_{\text{filtered}}[n-1].
\label{eq:ewma_shift}
\end{equation}
The first term in
\eqref{eq:ewma_shift} depends only on the current packet, so it can be
precomputed by stateless ALUs in earlier stages.
 
The second term, however, involves the previous sample
$x_{\text{filtered}}[n-1]$. This state must be stored in a register attached
to a stateful ALU in a single stage. A packet passes through each stage exactly once and cannot return to it. Hence, the state
update must be performed as a
single atomic register action. In this action, the
stateful ALU has to read $x_{\text{filtered}}[n-1]$, right-shift it by
$i$ bits, add the precomputed term, and write the result back. This requires a
fused read--shift--add--write operation, which current Tofino-style stateful
ALUs do not support.
Therefore, even with the restriction $\alpha = 1 - 2^{-i}$, the
EWMA filter cannot be realized on current programmable switches.

An extension of the stateful ALU removes this limitation. A recent proposal fuses the shift and the addition into a single atomic register update~\cite{yuan2022unlocking}.
This extension allows the entire read--shift--add--write sequence to
execute within one stage. Its hardware feasibility was verified through
synthesis, and the extended ALU was shown to meet timing. With such an extension,
\eqref{eq:ewma_shift} maps directly onto the pipeline. Earlier stages
precompute $x[n] - 2^{-i}x[n]$. Then, one stateful stage performs the
atomic update of $x_{\text{filtered}}[n]$.

\section{Conclusion}
In this paper, we proposed SPIN, a signal processing concept that executes signal processing directly on a programmable switch. Specifically, we implemented a binDCT-based block transform as a preprocessing stage for INML to validate the effectiveness and feasibility of our approach.
Experimental results demonstrated that filtering signals on a programmable switch improves classification performance in machine learning tasks. Furthermore, the evaluation confirmed the feasibility of implementing the proposed logic within the strict hardware resource constraints of a programmable switch.


\bibliographystyle{IEEEtran}
\bibliography{bibliography} 
\vfill\pagebreak

\end{document}